\documentclass[aps,floatfix,superscriptaddress,twocolumn]{revtex4}

\usepackage{graphicx}
\usepackage{amsmath,amssymb,amsfonts}
\usepackage{url,hyperref}
\usepackage{multirow}
\usepackage[version=4]{mhchem}

\usepackage{subcaption}
\usepackage{verbatim}
\usepackage{titlesec}
\usepackage{enumitem}
\usepackage{setspace}

\usepackage{color}
\usepackage[utf8]{inputenc}

\begin{document}


\title{Dynamic Local Structure in Caesium Lead Iodide: Spatial Correlation and Transient Domains}

\author{William Baldwin}
\email{wjb48@cam.ac.uk}
\affiliation{Department of Engineering, University of Cambridge, Cambridge CB2 1PZ, UK}

\author{Xia Liang}
\affiliation{Department of Materials, Imperial College London, London SW7 2AZ, UK}

\author{Johan Klarbring}
\affiliation{Department of Materials, Imperial College London, London SW7 2AZ, UK}
\affiliation{Department of Physics, Chemistry and Biology (IFM), Link\"{o}ping University, SE-581 83, Link\"{o}ping, Sweden}

\author{Milos Dubajic}
\affiliation{Department of Chemical Engineering and Biotechnology, University of Cambridge, CB3 0AS, UK}

\author{David Dell'Angelo}
\affiliation{CNR-IOM, Unità di Cagliari, 09042 Monserrato (CA), Italy}

\author{Christopher Sutton}
\affiliation{Department of Chemistry and Biochemistry, University of South Carolina, South Carolina 29208, United States}

\author{Claudia Caddeo}
\affiliation{CNR-IOM, Unità di Cagliari, 09042 Monserrato (CA), Italy}

\author{Samuel D. Stranks}
\affiliation{Department of Chemical Engineering and Biotechnology, University of Cambridge, CB3 0AS, UK}

\author{Alessandro Mattoni}
\affiliation{CNR-IOM, Unità di Cagliari, 09042 Monserrato (CA), Italy}

\author{Aron Walsh}
\affiliation{Department of Materials, Imperial College London, London SW7 2AZ, UK}

\author{G\'abor Cs\'anyi}
\affiliation{Department of Engineering, University of Cambridge, Cambridge CB2 1PZ, UK}


\begin{abstract}
Metal halide perovskites are multifunctional semiconductors with tunable structures and properties. They are highly dynamic crystals with complex octahedral tilting patterns and strongly anharmonic atomic behaviour. In the higher temperature, higher symmetry phases of these materials, several complex structural features have been observed. The local structure can differ greatly from the average structure and there is evidence that dynamic two-dimensional structures of correlated octahedral motion form. An understanding of the underlying complex atomistic dynamics is, however, still lacking. In this work, the local structure of the inorganic perovskite \ce{CsPbI3} is investigated using a new machine learning force field based on the atomic cluster expansion framework. Through analysis of the temporal and spatial correlation observed during large-scale simulations, we reveal that the low frequency motion of octahedral tilts implies a double-well effective potential landscape, even well into the cubic phase. Moreover, dynamic local regions of lower symmetry are present within both higher symmetry phases. These regions are planar and we report the length and timescales of the motion. Finally, we investigate and visualise the spatial arrangement of these features and their interactions, providing a comprehensive picture of local structure in the higher symmetry phases.
\end{abstract}

\maketitle

\date{\today}

\section{Introduction}

At finite temperatures metal halide perovskites exhibit complex and interesting structures. Methylammonium lead iodide (\ce{CH3NH3PbI3}) has been experimentally shown to exhibit static lattice twinning at room temperature\cite{Breternitz2020}, as well as a stress-sensitive ferroelastic domain structure \cite{Strelcov2017}. The average local strain of perovskite lattices has also been found to vary over large distances up to hundreds of nanometers or even micrometers\cite{Tennyson2019}. There are also open questions about how local distortions of the lattice couple to electronic degrees of freedom and to light\cite{Stranks2019, Wei2017}. 

~

Inorganic and organic perovskites have characteristic soft phonon modes which correspond to correlated octahedral tilting patterns. Along these modes, the potential energy surface can exhibit a shallow double well landscape, which gives rise to symmetry-breaking phase transitions \cite{Yang2017,Bechtel2018,Klarbring2018_pathways}. During dynamics, these modes are overdamped with very short lifetimes \cite{Songvilay2019, Fransson2022, Cohen2022, Carignano2017}. In the high temperature, high symmetry perovskite phases, population of these modes results in interesting phenomena. For instance, it has been experimentally shown in \ce{CsPbBr3} that excitations of these modes form a two-dimensional, planar structure \cite{Delaire2021}. Similar structures have been found in \ce{CH3NH3PbI3} using diffuse scattering experiments and classical molecular dynamics (MD) simulations \cite{weadock2023nature}.

~

In contrast to other Pb-I perovskites, less is known about \ce{CsPbI3}. The cubic phase of this material has a band gap that is well-suited for perovskite tandem solar cells\cite{Beal2016}. However, poor thermodynamic stability of the black perovskite phases at ambient conditions\cite{Yao2021} remains a significant challenge in photovoltaic applications. Since similar materials exhibit dynamic two-dimensional local structural features and local symmetry broken domains have already been reported in \ce{CsPbI3}\cite{bertolotti2017coherent}, the local structure is evidently complex. A detailed picture of the structural dynamics is therefore needed to tackle the stability of this material. 

~

To analyze the structural dynamics of this system on the necessary length and time scales, a fast and accurate simulation approach is required. To date, several empirical force fields with a fixed functional form have been developed for organic and inorganic perovskites. The MYP series of force fields 
\cite{Mattoni2015,Hata2017} has proved useful for modelling hybrid and inorganic perovskites, and was used to study several ionic properties such as the dielectric function of \ce{CH3NH3PbI3}\cite{Mattoni2020_dielectric} or the thermal conductivity in \ce{CsPbI3}\cite{Giri2022}.
The reactive force field ReaxFF and polarisable model AMEOBA have also been fitted to this material\cite{Pols2021, Rathnayake2020}. 
Both models are able to reproduce the phase transitions, but lack quantitative agreement for properties such as transition temperatures and ratios of lattice parameters. 
Alternatively, machine learning interatomic potentials (MLIPs) provide a flexible solution for accurate atomistic simulations\cite{Bartok2017, Musil2021, Butler2018, Deringer2021}. For material science, MLIPs are now becoming mature and have been applied to tackle scientific questions which would have been out of reach using {\em ab initio} or empirical methods\cite{Deringer2021silicon}. In the context of halide perovskites, several modern MLIP architectures have been demonstrated to date\cite{Jinnouchi2019, Fransson2023}.

~

In this work, we characterise the dynamic local structure of \ce{CsPbI3} using a new MLIP based on the atomic cluster expansion (ACE)\cite{Drautz2020} framework and large-scale simulations. Aiming to elucidate the three-dimensional structure and dynamics of \ce{CsPbI3}, we analyse spatial and temporal correlations of octahedral tilting angles for the three perovskite phases of the material. By investigating the temporal correlation of octahedral tilting in section \ref{sec:temporal}, we reveal that a double well effective potential is active in the cubic phase. In section \ref{sec:spatial}, we confirm that planar correlation structures observed in other materials also form in \ce{CsPbI3}. Furthermore, it is shown that the three dimensional octahedral tilting pattern of these local lower symmetry regions is inherited from the lower temperature phases. In sections \ref{sec:cubic} and \ref{sec:tetragonal}, the spatial arrangement and interactions of these lower symmetry regions is investigated. A detailed picture of the nature of the cubic and tetragonal phases is then presented. While planar regions of correlated octahedral motion exist, these alone do not fully describe the local structure. The spatial arrangement and overlapping nature of these features is needed to properly understand these symmetry-broken domains.

\section{Results and Discussion}

\subsection{Machine Learning Potential Validation}

\begin{figure*}[t]
  \begin{minipage}{0.49\textwidth}
    \centering
    \includegraphics[width=1.0\linewidth]{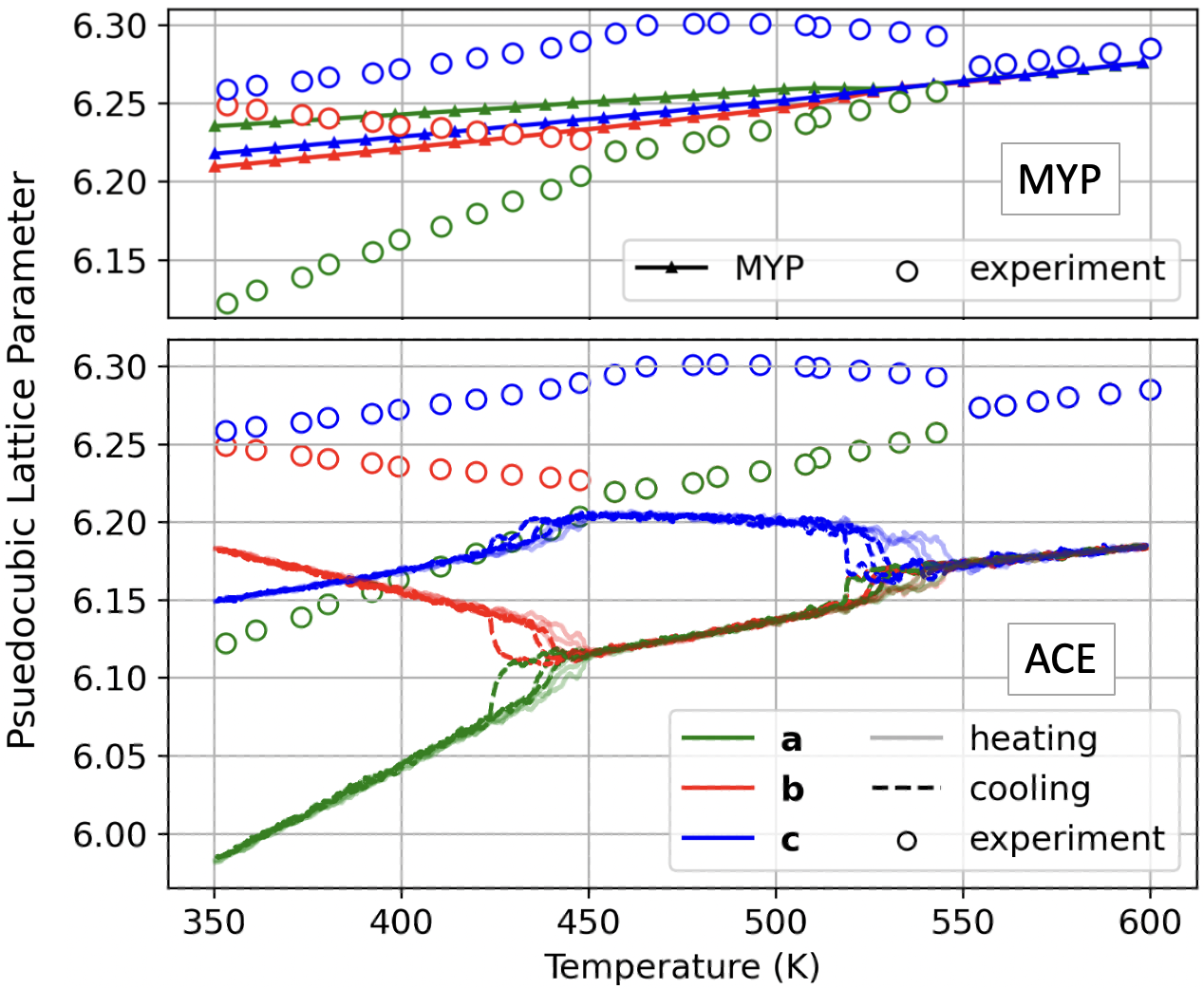}
    \captionof{figure}{Evolution of the pseudocubic lattice constants of \ce{CsPbI3} for the ACE potential and the MYP empirical force. These simulations were performed using a 13720 atom simulation cell cooling over 4 ns. For ACE, three traces are shown for both heating and cooling. Experimental data reproduced with permission from Even and coworkers.\cite{Even2018} \label{fig:lattice_param_variation}}
  \end{minipage}
  \hfill
  \begin{minipage}{0.49\textwidth}
    \centering
    \setlength\tabcolsep{0pt}
    \begin{tabular*}{\linewidth}{@{\extracolsep{\fill}} cccc}
\hline
\multicolumn{1}{l}{} & \multicolumn{3}{c}{Temperature (K)} \\ 
\multicolumn{1}{l}{} & 325 & 510 & 645 \\ \hline
a (\AA)& 5.95 & 6.14 & 6.19 \\  
a rescaled & 6.05 (6.095) & 6.24 (6.214) & 6.29 (6.297) \\ \hline
b (\AA) & 6.20 & 6.14 & 6.19 \\
b rescaled & 6.30 (6.259) & 6.24 (6.241) & 6.29 (6.297) \\ \hline
c (\AA) & 6.14 & 6.20 & 6.19 \\
c rescaled & 6.24 (6.250) & 6.30 (6.299) & 6.29 (6.297) \\ \hline
$\alpha$ (degrees) & 12.4 (11.5) & 9.0 (8.6) & 0.0 (0.0) \\
$\delta$ (degrees) & 8.0 (7.0) & 0.0 (0.0) & 0.0 (0.0) \\ \hline
\end{tabular*}
  \captionof{table}{Finite temperature structural predictions for \ce{CsPbI3} compared to experimental data from Even and coworkers \cite{Even2018}. Experimental values are shown in parentheses.\label{table:finite_temperature_structure}}
\end{minipage}
\end{figure*}

A machine-learned interatomic potential based on the ACE framework has been trained for \ce{CsPbI3} as described in Section \ref{section:methods}. 
\ce{CsPbI3} has four distinct phases which appear between room temperature and 600 K. There are three perovskite phases that form on cooling from 600 K to 350 K: a cubic phase ($\alpha$), a tetragonal ($\beta$) and an orthorhombic ($\gamma$). 
At room temperature, the system transitions to a non-perovskite edge-sharing polymorph ($\delta$) \cite{Even2018}. 

~

Our model has been trained to describe all of these phases and validated by predicting experimental observables. In particular, we computed the variation of pseudocubic lattice parameters predicted by the model as a function of temperature. This was done by running a simulation of 13720 atoms (14$^3$ pseudo cubic unit cells) in the NPT ensemble with a slowly varying temperature. The average pseudocubic lattice parameters can then be computed from the trajectory. \textbf{Figure \ref{fig:lattice_param_variation}} shows the predicted  pseudocubic lattice parameter variation compared to experimental data obtained by \textit{in situ} synchrotron X-ray diffraction provided by Even and coworkers \cite{Even2018}. The prediction of the MYP empirical force field \cite{Giri2022} is also shown for comparison because force fields of this type have successfully been used to study organic perovskites. As it can be seen from Figure \ref{fig:lattice_param_variation}, the model is successful in predicting the overall experimental trends. There is an offset as the model under-predicts the lattice parameters by $1.6\%$. This is a feature of the density functional theory (DFT) reference data used to train the model, which utilised the local density approximation (LDA). In this 13720 atom box and with a heating and cooling rate of 62.5 K ns$^{-1}$, some variability in the phase transition temperature is observed, but the nature of the transitions is correct. For discussion of finite time effects in such phase transitions, readers are referred to recent work by Fransson and coworkers\cite{Fransson2023}.

~

In \textbf{Table \ref{table:finite_temperature_structure}} we show a numerical comparison of the experimental structural characteristics of the pseudocubic unit cell and those computed by simulations of a 1000 atom system in the NPT ensemble and averaging the atomic positions over time. We also report the re-scaled lattice parameters corresponding to a uniform $5\%$ re-scaling of the volume. Overall, the model performs well, capturing the ratios of lattice constants and tilting angles accurately. Readers are referred to the supplementary material for further MLIP validation. The fact that the MYP force field is unable to capture the above structural features illustrates the need for machine learning potentials for studying this material.

\subsection{Multiple Timescales of Octahedral Motion}
\label{sec:temporal}

\begin{figure}
    \centering
    \includegraphics[width=0.9\linewidth]{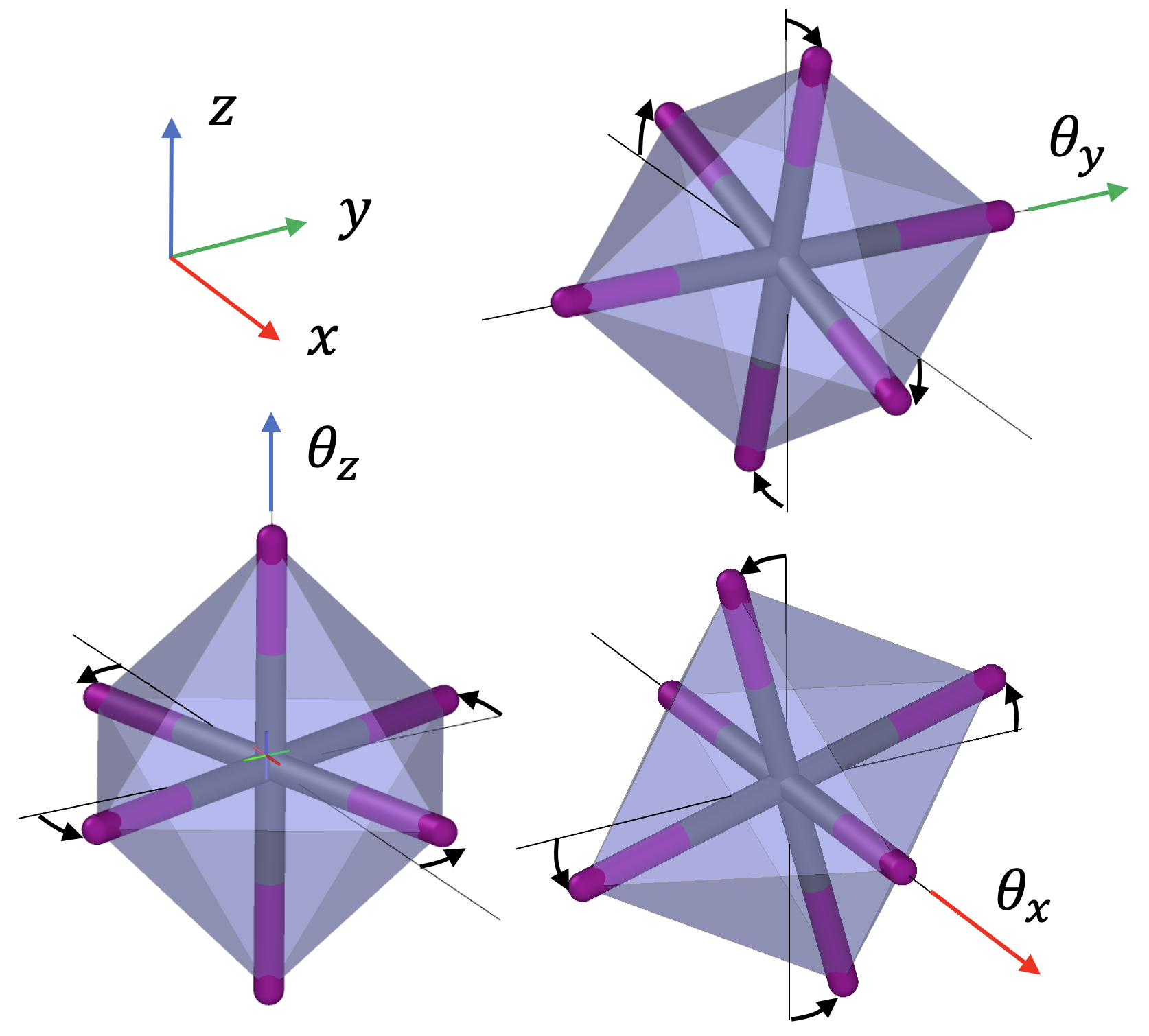}
    \caption{Visualisations of the \ce{PbI6} octahedral tilting angles used in this study. In the limit of small rotations, the Euler angles which map a rotated octahedron to one aligned with the coordinate axes correspond to the rotations shown here.}
    \label{fig:angles_definition}
\end{figure}

Structural phase transitions of the general ABX$_3$ perovskite can occur through three types of distortions \cite{howard_structures_2005}: (1) octahedral deformations, (2) B-cation displacements, and (3) octahedral rotations. 
The latter commonly act as an order parameter that drives structural phase transitions in halide perovskites. Consequently, our focus will be primarily on these octahedral rotations.
We define three tilting angles for each \ce{PbI6} octahedron. This is done by first finding a best fit regular octahedron to a particular Pb site and then taking the Euler angles which map this octahedron to a perfect one in the laboratory reference frame.
One can choose a convention to order the Euler angles such that in the limit of small rotations, the three angles correspond to signed tilting about the three coordinate axes. This is illustrated in \textbf{Figure \ref{fig:angles_definition}}. Further detail about the tilting angle computation is given in the supplementary material. We introduce the following notation to refer to tilting angles at a given time and position: 
\begin{align*}
    \theta_i(t;n_x,n_y,n_z)
\end{align*}
where $i$ is the direction of tilt ($i=x$ refers to tilting around the $x$-axis), $t$ is the given time and $(n_x,n_y,n_z)$ are integer coordinates indexing the pseudocubic lattice site of the octahedron. All angles are reported in degrees.

~

\begin{figure*}[htb]
    \centering
    \includegraphics[width=0.8\textwidth]{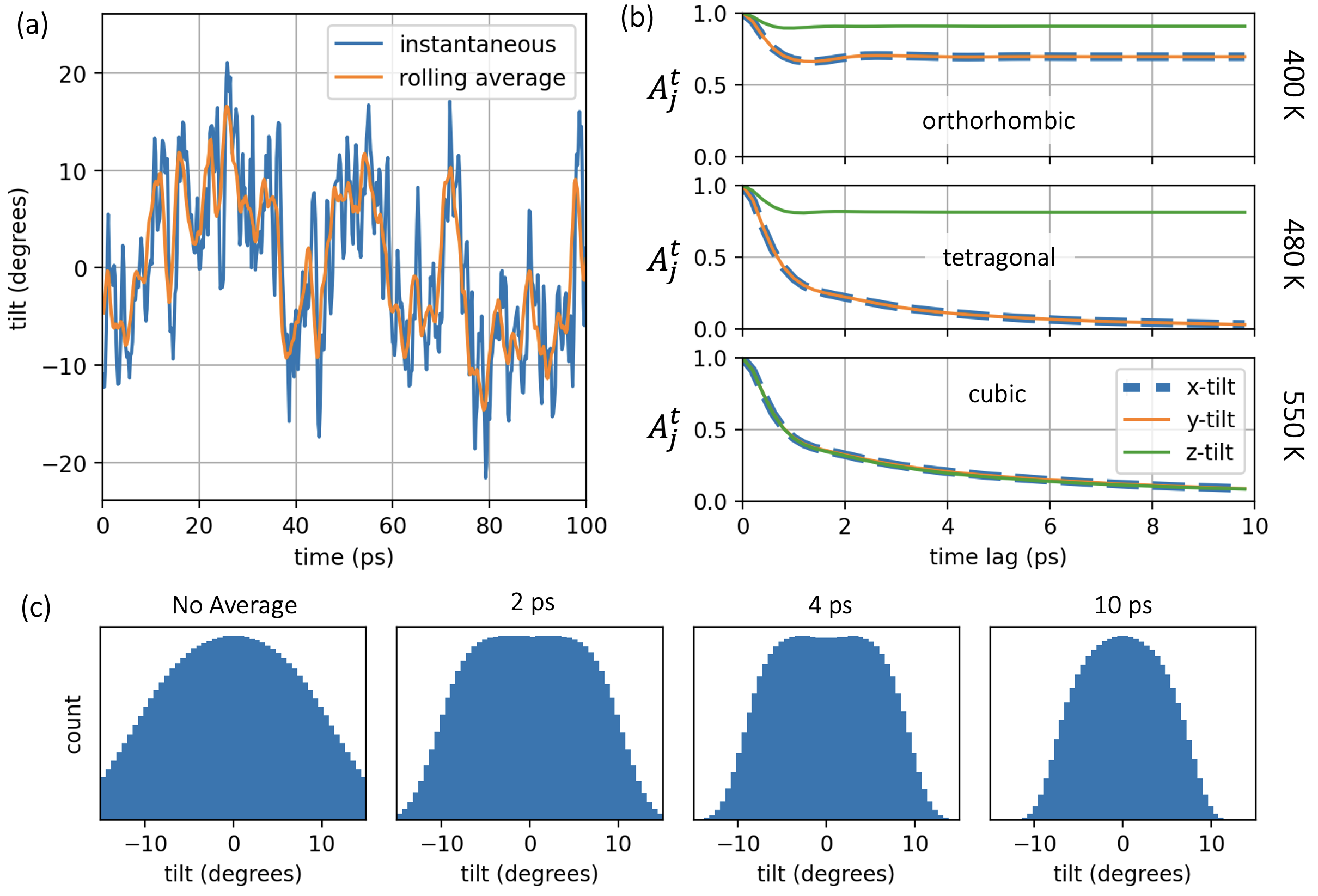}
    \caption{Temporal correlation structure of \ce{PbI6} octahedral tilts in \ce{CsPbI3}. (a): The $x$-direction tilt ($\theta_x$) of a single octahedron over the course of 100 ps at 550 K with a 200 fs sampling interval, and after applying a 2 ps rolling average. (b): Temporal autocorrelation of all octahedra in the three perovskite phases. (c): Histograms of instantaneous $\theta_x$ values at 550 K, as well as the same tilt after performing a rolling average over various timescales.}
    \label{fig:temporal_struc_compound}
\end{figure*}

To understand the timescales of tilting dynamics, we have examined the temporal autocorrelation function of local octahedral tilts. This describes how a single tilting angle is correlated with itself at a future time. With the above notation, the temporal autocorrelation function is defined as
\begin{align}
    A^{\text{temporal}}_i(\tau) = \frac{1}{\mathcal{N}} \ \mathbb{E}_{t,\mathbf{n}} [ \ \theta_i(t;n_x,n_y,n_z) \ \theta_i(t+\tau;n_x,n_y,n_z) \ ]
    \label{eq:TACF}
\end{align}
where $\mathbb{E}_{t,\mathbf{n}}$ denotes the expectation over time and lattice sites and $\mathcal{N}$ is such that $A^{\text{temporal}}_i(0)=1$. \textbf{Figure \ref{fig:temporal_struc_compound}b} shows the three components of $A^{\text{temporal}}_i$ as a function of temperature. This data was collected from simulations of 69120 atoms ($24^3$ pseudo cubic unit cells).

~

The asymptotes of the correlation functions reflect the Glazer notation\cite{glazer_classification_1972} of each phase: In the cubic phase at 550 K, all three correlations decay to zero with time since all tilts are disordered with zero mean over long time scales ($a^0 a^0 a^0$). In the tetragonal phase, which has a tilt-configuration of $a^0 a^0 c^+$, only $\theta_z$ has a non-zero mean ($c^+$). This is reflected in the autocorrelation function at 480 K where only the $A_z(\tau)$ converges to a non-zero value. The same holds at 400 K in the orthorhombic  phase, where the correlation of all three tilts converge to non-zero values and the $\theta_x$ and $\theta_y$ correlations are identical, in agreement with its $a^- a^- c^+$ tilt configuration.  

~

At 400 K, the autocorrelation for $\theta_x$ and $\theta_y$ show evidence of a damped oscillatory mode with a time period on the order of 2 ps. In fact, as shown in the supplementary information, the $A^t_i$ for the ordered tilts can be fitted accurately to damped harmonic oscillators. For the disordered tilts, at 480 K and 550 K, we find an initial relaxation on the same timescale, followed by a longer decay component. To explain the shape of these correlation functions we examine the tilt of a single octahedron tilt as a function of time. In \textbf{Figure \ref{fig:temporal_struc_compound}a}, the $\theta_x$ value of a single octahedron at 550K is plotted over the course of 100 ps. Also shown is the rolling average of this trajectory. A square window is used to perform the average - the angle at a given time is replaced by the average value over the previous 2 ps. One can see evidence of a long time scale wandering motion of the averaged value, as well as high-frequency oscillations around the average. This behaviour is reminiscent of analysis by Fransson \cite{Fransson2022} of octahedral tilting phonon modes in \ce{CsPbBr3}.
 
~

We have explored this further by examining the histogram of tilting angles after first performing rolling averages. \textbf{Figure \ref{fig:temporal_struc_compound}c} shows the histogram of $\theta_x$ for both the raw data and after applying a rolling average of the tilting angles with a range of averaging timescales. For this analysis a square window was used to compute the rolling averages, however equivalent results are found by using a first-order Butterworth low pass filter\cite{butterworth} to remove the high-frequency components. In the instantaneous (without time averaging) picture, we find that the tilts are distributed in a unimodal distribution around zero. Naively, this would seem to indicate that the octahedra simply oscillate around their cubic, zero-tilt, reference positions, and that no local structure is present in the material.  However, on averaging over a 1-10 ps timescale, a flat topped or double peaked distribution emerges, indicating that on this timescale the effective potential for octahedral rotation in fact has minima at non-zero tilt angles. 

~

An explanation for this behaviour is that the average tilt is hopping between two minima with a characteristic time on the order of 5 ps. Superposed on this average value are high-frequency oscillations around the current minimum. If the high-frequency oscillations give a Gaussian distribution about the current mean, then the total distribution would be the sum of these two such Gaussians at the two minima. For broad Gaussians, this sum would merge the two peaks, giving a unimodal instantaneous tilt distribution. If averages of the order of 4 ps are performed, however, these deviations from the mean are reduced, revealing the two minima. If an average is performed over 10 or more picoseconds, then the mean also jumps between both wells and the average converges to zero. Given that the double peak structure disappears when averaging over 10 ps, we can conclude that the characteristic hopping time is less than 10 ps. This can be compared to simulations performed by Fransson and coworkers \cite{Fransson2022}. They performed similar ML-MD simulations of \ce{CsPbBr3}, but instead decomposed atomic motions into supercell phonon modes. They reported that close to the upper transition in \ce{CsPbBr3}, the autocorrelation of the phonon mode component associated with in-phase tilting had a characteristic decay time of 5.22 ps. 
With this interpretation, the temporal autocorrelation functions for the tetragonal and cubic structures (at 480 K and 550 K, respectively in Figure \ref{fig:temporal_struc_compound}b) can be viewed as the superposition of the decorrelation of the high-frequency oscillatory behaviour (as revealed in the orthorhombic at 400 K) and the longer timescale average tilt decorrelation. 

~

The temporal correlation structure of the tilts in cubic \ce{CsPbI3} indicates that multiple timescales are present. Performing temporal averaging of the tilts suggests that high frequency oscillations are masking an underlying double well effective potential. We can also conclude the characteristic hopping time of the mean value is between 4 and 10 ps. This effect has been shown in the cubic phase at 550 K, close to the cubic to tetragonal transition temperature of this potential of 533 K. 

\subsection{Spatial Correlation Structure}
\label{sec:spatial}

\begin{figure*}[t]
    \centering
    \includegraphics[width=0.9\textwidth]{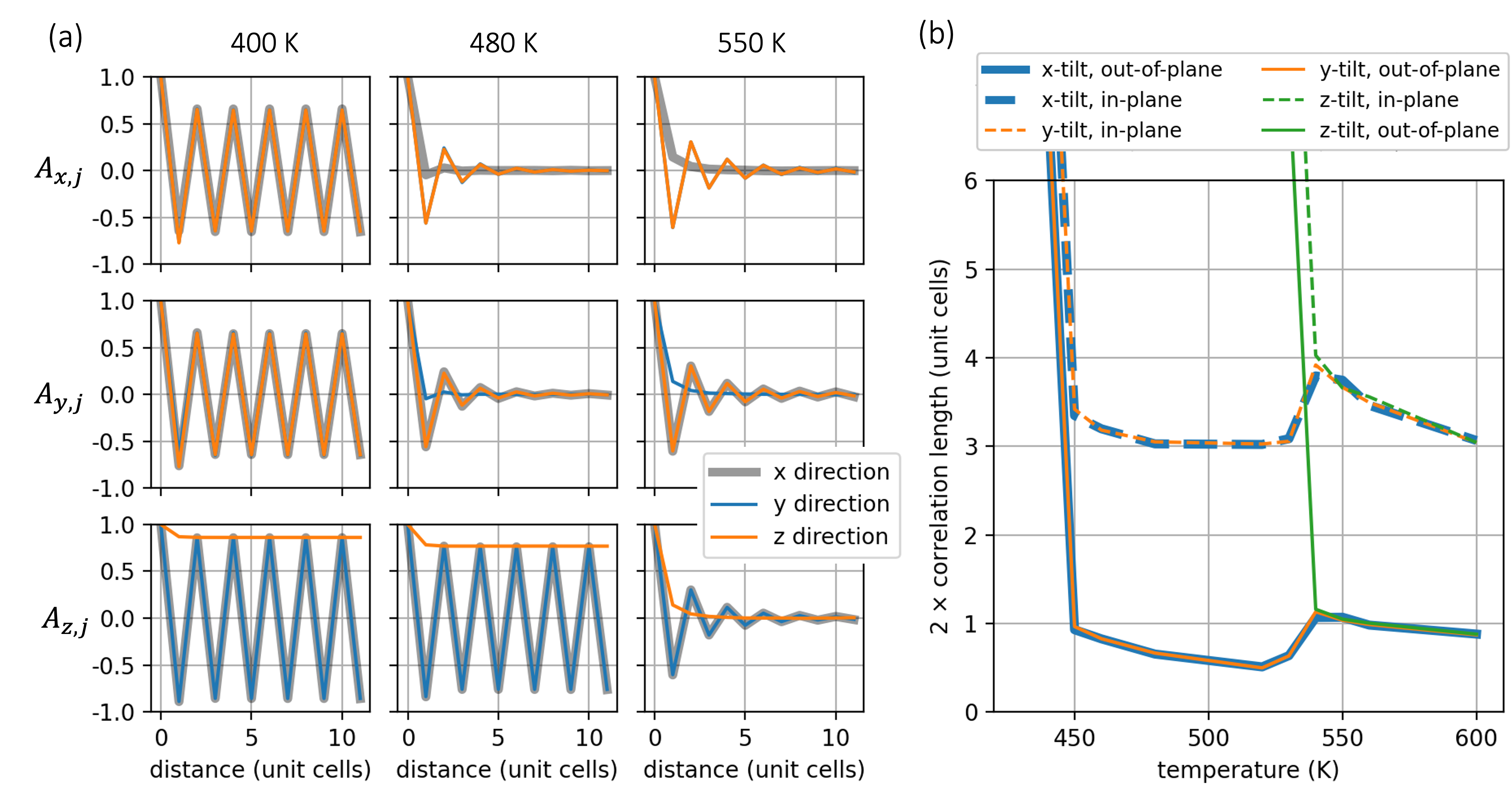}
    \caption{Spatial correlation structure  of \ce{PbI6} octahedral tilts in \ce{CsPbI3} with temperature. (a): Spatial autocorrelation function for all tilting directions in the three perovskite phases. (b): Trends in correlation length as a function of temperature.}
    \label{fig:spatial_corr_compound}
\end{figure*}

The spatial structure of tilting dynamics can be studied using similar methods. In analogy to Equation \ref{eq:TACF}, we can define the spatial autocorrelation function as:
\begin{align*}
    A^{\text{spatial}}_{i,x}(d) &= \frac{1}{\mathcal{N}} \ \mathbb{E}_{t,\mathbf{n}} [ \ \theta_i(t;n_x,n_y,n_z) \ \theta_i(t;n_x+d,n_y,n_z) \ ] \\
     A^{\text{spatial}}_{i,y}(d) &= \frac{1}{\mathcal{N}} \ \mathbb{E}_{t,\mathbf{n}} [ \ \theta_i(t;n_x,n_y,n_z) \ \theta_i(t;n_x,n_y+d,n_z) \ ] \\
      A^{\text{spatial}}_{i,z}(d) &= \frac{1}{\mathcal{N}} \ \mathbb{E}_{t,\mathbf{n}} [ \ \theta_i(t;n_x,n_y,n_z) \ \theta_i(t;n_x,n_y,n_z+d) \ ]
\end{align*}
The function $A^{\text{spatial}}_{i,j}$ has nine components since there are three tilting angles, and we can take the autocorrelation along three spatial directions. For example, $A_{x,y}(d)$ describes how the correlation of $\theta_x$ decays as one moves away in the $y$-direction. \textbf{Figure \ref{fig:spatial_corr_compound}a} shows the spatial autocorrelation function for a range of temperatures spanning the three perovskite phases. 

~

At 550 K, the spatial correlation function shows that each tilting direction exhibits anti-phase correlation along two spatial directions, which decays over the course of about 5 unit cells, as well as a weak in-phase correlation in the third direction which decays within 1 unit cell. Importantly, the spatial correlation length of $\theta_i$ is short in the $i$-direction, but long in the perpendicular directions. This can be interpreted as a two dimensional, planar correlation structure and is a result of the corner connectivity of the perovskite octahedral network \cite{glazer_classification_1972}. All three octahedral tilting components exhibit their own planar correlation structure: $\theta_x$ is correlated in the $y$--$z$ plane, $\theta_y$ is correlated in the $x$--$z$ plane and $\theta_z$ is correlated in the $x$--$y$ plane.

~

As the temperature decreases and the material undergoes a phase transition, certain octahedral tilts become `locked' as a result of symmetry breaking. This corresponds to the spatial autocorrelation function not approaching zero at large distances. According to the Glazer notation, the average structure of the tetragonal phase of \ce{CsPbI3} can be denoted as $a^0 a^0 c^+$. This is captured in our simulations at 480 K. The correlations of $\theta_x$ and $\theta_y$ across all three spatial directions have an asymptote of zero ($a^0 a^0$) while the correlation of $\theta_z$ does not decay to zero and has in-phase correlation in the $z$-direction, corresponding to the $c^+$ in the Glazer notation. Due to the rigid coupling of the halide ions, $\theta_z$ always has anti-phase correlation in the in-plane directions ($x$ and $y$)\cite{glazer_classification_1972}.

~

The autocorrelation functions in Figure \ref{fig:spatial_corr_compound}a reveal a surprising feature of the material. In the cubic phase, the out-of-plane correlation of $\theta_i$ is in-phase. This can be compared to the sign of the out-of-plane correlation of $\theta_z$ (the `locked' tilt) in the tetragonal phase, which is also in-phase. The dynamic correlations in the cubic phase therefore share the same three dimensional tilting pattern as the locked tilting angle in the tetragonal phase. A similar story holds for the dynamic correlations in the tetragonal phase. Here, the out-of-plane correlations for $\theta_x$ and $\theta_y$ are very small, but negative. This mirrors the pattern in the orthorhombic phase. The result is that the dynamic correlations in both phases have the same Glazer tilting pattern as the `locked' tilt in phase immediately below. The structure of the lower symmetry phases is preserved locally in the dynamic structure of higher symmetry phases.

~

Finally, one can also look at the variation in correlation length with temperature, which is shown in \textbf{Figure \ref{fig:spatial_corr_compound}b}. The spatial correlation length was calculated by fitting an exponential decay ($e^{-d/\lambda}$) to the absolute values of the autocorrelation function. In Figure \ref{fig:spatial_corr_compound}b, we have plotted $2\lambda$ which corresponds to the thickness or diameter of the correlated regions. One can see that the characteristic length scales in the tetragonal and cubic phases increase slightly as the temperature lowers towards each phase transition.
In the context of instantaneous and time-averaged structure, one finds the same qualitative results for the spatial autocorrelation function when first performing time averaging of the tilting angles over 2 ps. 

\subsection{Onset of the Cubic Phase}
\label{sec:cubic}

So far, we have shown that both the tetragonal and orthorhombic phases have a planar structure of tilting correlations. Furthermore, in both of these phases, the local transient tilting correlation is the same as the global pattern in the next lower symmetry phase. The natural question is whether these correlations translate to local domains of tetragonal material within the cubic phase, or orthorhombic material in the tetragonal.
If this is the case, one can also ask how these domains are arranged spatially. To properly characterise the cubic phase, we have assessed the correlation between different tilts at a single site, and visualised the three dimensional structure of the correlated tilts in the material. 

~

\begin{figure*}[t]
  \includegraphics[width=0.9\textwidth]{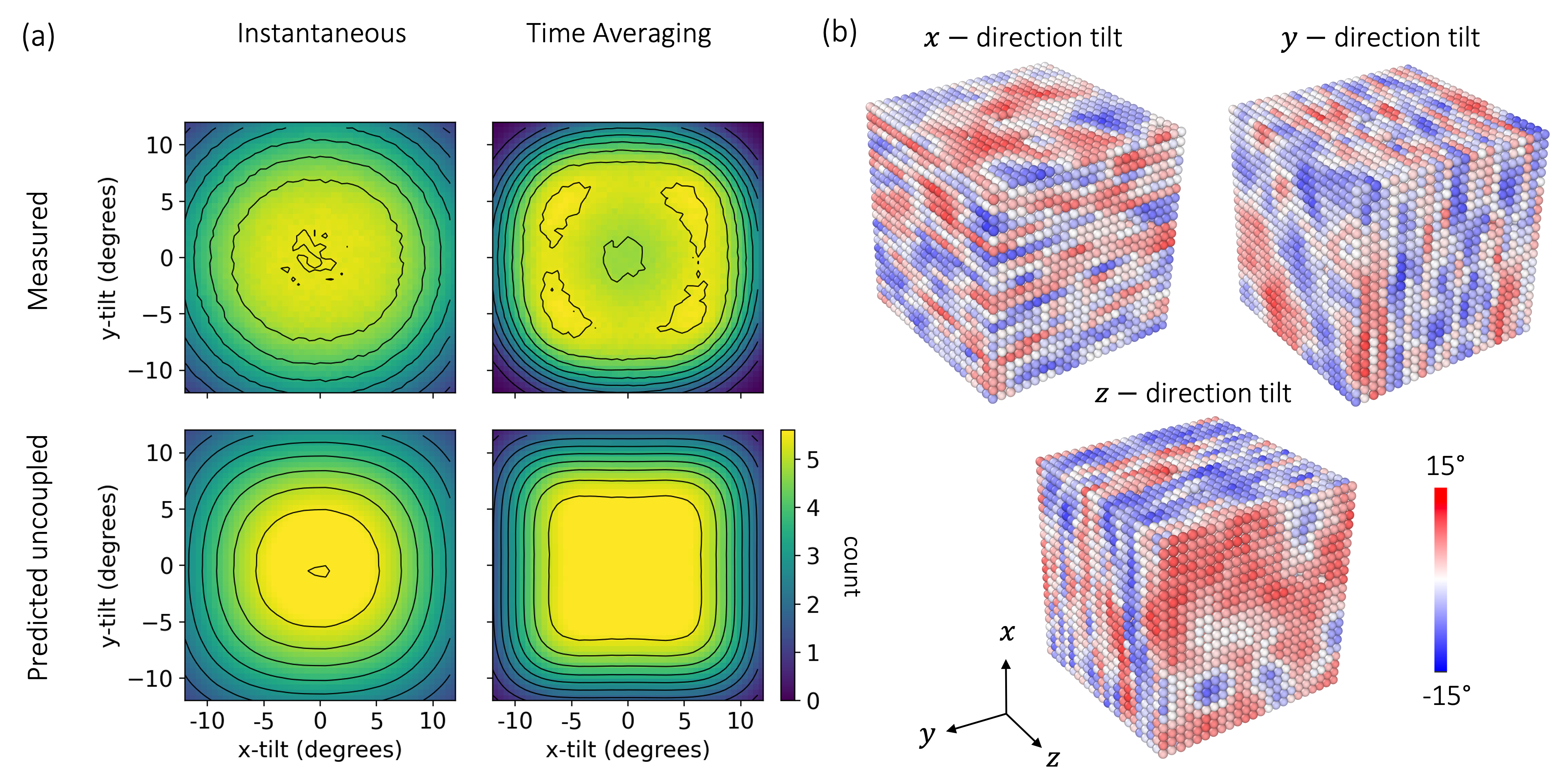}
  \caption{Nature of the cubic perovskite phase of \ce{CsPbI3}. (a): Heat maps of $\theta_x$ and $\theta_y$ in the cubic phase at 550 K. The top row is the observed data, and the bottom row is the predicted distribution assuming that the two tilts are independent random variables. We have shown the results for both instantaneous tilts and for a rolling average of 2 ps. The scale (count) is arbitrary, and has been re-scaled for ease of interpretation (b): Visualisation of the time averaged tilting structure. In each plot, the lead atoms are drawn with the colour corresponding to one of the three tilting angles. Visualisations were produced using \textsc{Ovito}.}
  \label{fig:cubic}
\end{figure*}

Firstly, one can generalise the histograms in Figure \ref{fig:temporal_struc_compound}c to show the joint distribution of multiple tilting directions. \textbf{Figure \ref{fig:cubic}a} shows a heat map of $\theta_x$ and $\theta_y$ for all octahedra at 550 K, close to the tetragonal transition temperature of this machine learned potential of 533 K. Since the cubic phase is isotropic, all three tilting angles have the same distribution. Similarly, only the heat map of $\theta_x$ and $\theta_y$ is needed, since the other pairings will be identical. Heat maps of both instantaneous tilts, and the same quantity after applying a 2 ps rolling average, are shown. One can see that instantaneously the joint distribution is unimodal. Furthermore, there is no correlation between the different tilting angles - they are independent random variables. We can see this by comparing to the predicted heat map assuming that the two tilting angles are independent variables. This was computed as the product of the marginal distributions of each tilt, as discussed in the supplementary information. The two heat maps are very similar, implying that excitations of these tilting modes do not interact.

~

On the 2 ps timescale, the story is qualitatively different. Instead of a unimodal distribution, we find a multimodal distribution with a significant drop in intensity at the center and four faint peaks. The modes of the measured distribution are not on the axes, but instead in the four quadrants. This suggests that most octahedra have multiple large non-zero tilts. The interpretation - in terms of local domains of tilted octahedra - is that if planar regions of correlated tilting angle form, then different regions with different orientations typically overlap, rather than being mutually exclusive. Furthermore, when comparing to the predicted distribution given that the tilts are independent, it is clear that on this time scale the two variables are not independent. While the two plots share many features, there is a substantial decrease in intensity at the center indicating a strong tendency to avoid having both tilting angles equal to zero. In summary, Figure \ref{fig:cubic}a shows that on the 2 ps timescale, most octahedra have multiple large nonzero tilts, and the excitations of different tilting mode directions are coupled. 

~

This behaviour has been presented close to the tetragonal transition temperature. Similar analysis at higher temperatures revealed that multimodality in the 2D and 3D joint tilt distributions persist for many tens of kelvin above the transition temperature, as shown in the supplementary information.
In particular, even at 600K (77 K above the transition temperature of this potential) there is still a strong tendency to avoid having all three tilting angles equal to zero, clearly demonstrating that local structure of lower symmetry is present even deep into the cubic phase.

~

Finally, one can visualise this behaviour. This has been done by taking frames from the simulations used in the above analysis and colouring the octahedra according to a particular angle of tilt. To visually appreciate the size and shape of homogeneous regions, it is necessary to apply a mask to the tilting angles. This is because a homogeneous region of correlated $\theta_x$ in the cubic phase exhibits anti-phase correlation in two axes, and in-phase correlation in the third. More precisely, we can see from Figure \ref{fig:spatial_corr_compound}a that the correlations have the same parity pattern as the average structure in the tetragonal phase. In Glazer notation, this is $a^0 a^0 c^+$. Therefore, we can apply the following operation to $\theta_x$ such that homogeneous regions with the tilting pattern of the tetragonal phase are rendered the same colour:
\begin{align*}
    \theta_x(t;n_x,n_y,n_z)  \quad &\rightarrow \quad (-1)^{n_y} (-1)^{n_z} \theta_x(t;n_x,n_y,n_z)
\end{align*}
A similar operation can be done to the other tilting directions:
\begin{align*}
    \theta_y(t;n_x,n_y,n_z) \quad &\rightarrow \quad (-1)^{n_x} (-1)^{n_z} \theta_y(t;n_x,n_y,n_z) \\
    \theta_z(t;n_x,n_y,n_z) \quad &\rightarrow \quad (-1)^{n_x} (-1)^{n_y} \theta_z(t;n_x,n_y,n_z)
\end{align*}
A single frame of the trajectory, after performing time averaging over 2 ps, has been processed in this way and is shown in \textbf{Figure \ref{fig:cubic}b}. The three subplots correspond to colouring the octahedra according to the $x$-, $y$- and $z$-tilts respectively.

~

There is a clear planar structure in the patterns for each tilting direction. The normal of the planes is aligned with the tilting direction, as suggested by the spatial correlation functions. One can also see that areas which are largely homogeneous with large tilting angle when looking at one tilting direction intersect with equivalent regions in other plots. The tilted regions form and overlap with one another - in the same place, at the same time. Supplementary video 1 shows these visualisations over 400 ps of simulation time. When making the same visualisations without the time averaging, a similar planar structure is observed. 

~

On this basis of these results, interpreting the cubic phase as a dynamic average of the tetragonal phase is difficult. 
The planar structures do share the correct tilting pattern, however, they form together, with different orientations, overlapping one another. In fact, the formation of these regions is coupled and they are slightly more likely to form together, rather than being mutually exclusive. This leads to a local structure which is not characteristic of a specific orientation of the tetragonal phase. Instead, most regions appear to be at the intersection of multiple planes of correlated tilts, and thus have an even lower local symmetry. 

\subsection{Nature of the Tetragonal Phase}
\label{sec:tetragonal}

\begin{figure*}[t]
  \includegraphics[width=0.9\textwidth]{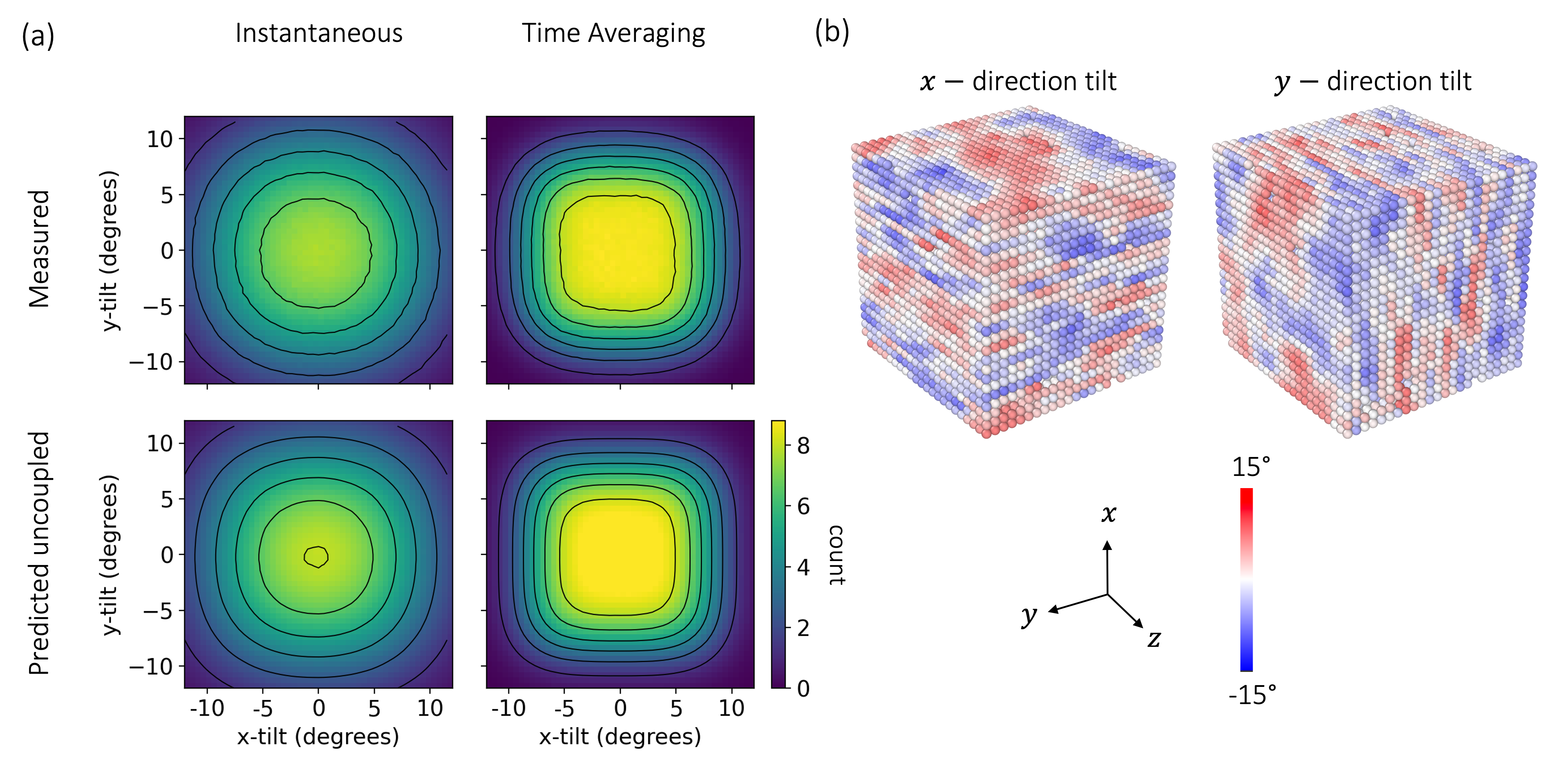}
  \caption{Nature of the tetragonal phase of \ce{CsPbI3}. (a): Heat maps of the x- and y-direction tilts in the tetragonal phase at 450 K. The scale (count) is arbitrary and has been re-scaled for ease of interpretation. (b): Visualisation of the time averaged tilting structure. In each plot, the lead atoms are drawn with the colour corresponding to one of the three tilting angles. Visualisations were produced using \textsc{Ovito}.}
  \label{fig:tetragonal}
\end{figure*}

The same analysis can be applied to the tetragonal phase, as shown in \textbf{Figure \ref{fig:tetragonal}}. Here, the $z$-direction tilt is globally locked. Therefore even though the structure is not isotropic, only the heat map of the two disordered tilting angles is shown. Qualitatively different behaviour is found, in that the tilting heat maps for both the instantaneous and time-averaged data show excellent agreement between the measured and the predicted data given independent variables. Correlated planes in the tetragonal phase can robustly be viewed as uncoupled, non-interacting excitations on both timescales. Furthermore, the time averaged picture shows almost no splitting of the maximum. The same results are found at higher temperatures within the tetragonal range. 

~

An equivalent visualisation is also shown. A different filter is applied to the tilting angles, since the dynamic correlations in this phase share the tilting pattern of the orthorhombic material, with anti-phase out-of-plane correlation (as in Figure \ref{fig:spatial_corr_compound}). Homogeneous regions are therefore rendered the same colour by the following mapping:
\begin{align*}
    \theta_x(t;n_x,n_y,n_z)  \quad &\rightarrow \quad (-1)^{n_x} (-1)^{n_y} (-1)^{n_z} \theta_x(t;n_x,n_y,n_z) \\
    \theta_y(t;n_x,n_y,n_z)  \quad &\rightarrow \quad (-1)^{n_x} (-1)^{n_y} (-1)^{n_z} \theta_y(t;n_x,n_y,n_z)
\end{align*}
Again, one can see the planar correlation in local tilting in the two disordered tilts. Supplementary video 2 shows these visualisations over the full trajectory. 

The local structure of the tetragonal phase is, again, difficult to interpret in terms of local symmetry broken domains. While each tilting angle exhibits planar correlated structures, with the 3D tilting pattern of the orthorhombic material, the dynamics of the two disordered tilting angles are independent. The local structure is a superposition of these separate excitations.

\section{Conclusions}

Extensive large simulations using an accurate machine-learned interatomic potential have been used to study the local structure of \ce{CsPbI3}. Our study confirms that the local structure in the halide perovskite is dynamic in both space and time.
We have shown through temporal averaging of the octahedral tilts that a double well behaviour can still be found in the cubic phase. The timescale associated with the hopping motion is in the 5--10 ps range. 

~

Analysis of the spatial autocorrelation function has revealed a planar correlation structure in the tetragonal and cubic phases. Furthermore, the local transient regions in these phases show a three-dimensional tilting pattern which matches that of the Glazer tilting pattern next lower temperature phase. Visualisations reveal that these correlations do indeed correspond to large planar regions of similarly tilted octahedra. Analysis of the joint distribution of tilts then reveals that these planar regions instantaneously appear to be uncoupled excitations. On the other hand, when removing higher frequency oscillations through the same time averaging procedure, one finds some correlation in these features. In the cubic phase, the structure is composed of dynamic, preferentially overlapping two dimensional tilted regions. In the tetragonal phase, the behaviour is qualitatively different, with the planar tilting excitations being independent both instantaneously and on longer timescales. Questions remain about how specific this behaviour is to \ce{CsPbI3} and we hope this will be investigated in the future.

~

The accuracy and efficiency of modern machine learning potentials have made this work possible. Due to the large correlation lengths and long timescales, these simulations would have been impossible using first-principles methods. At the same time, empirical force fields which perform well for organic systems are unable to quantitatively describe the nature of systems such as this inorganic perovskite. Systematically convergable machine learning frameworks such as the atomic cluster expansion are therefore creating new avenues of potential research in materials science. 

\section{Acknowledgements}

WB thanks the US AFRL for funding the work for this project through grant FA8655-21-1-7010.
JK acknowledges support from the Swedish Research Council (VR) program 2021-00486. 
We are also grateful for use of the ARCHER2 UK National Supercomputing Service (\url{http://www.archer2.ac.uk}) which is funded by EPSRC via our membership of the UK Car-Parrinello Consortium (grant reference EP/P022065/1), as well as our membership of the UK's HEC Materials Chemistry Consortium (grant reference EP/X035859/1). 
AM and CC acknowledge financial support from CNR for Short Term Mobility 2020 prot. 53767, and ICSC – Centro Nazionale di Ricerca in High Performance Computing, Big Data and Quantum Computing, funded by European Union – NextGenerationEU - PNRR, and computational support from CINECA under ISCRA initiative. 
The authors thank the Leverhulme Trust (RPG-2021-191) for funding.
The authors also thank Volker Blum for help and advice on the electronic structure calculations.

\section{Conflicts of Interests}

The authors declare no conflicts of interest.

\section{Data Availability Statement}

The machine learning potential, training and test datasets and example scripts on how to run the potential in LAMMPS are available at \url{https://github.com/WillBaldwin0/cspbi3_mlip_docker}. A docker image with a working LAMMPS installation and example scripts is also available through the same link. The trajectory files of all simulations that we have analysed will be uploaded and made available on publication. 

\section{Methods Section}
\label{section:methods}

\subsubsection*{Atomic Cluster Expansion}

The machine learning model in this work is based on the Atomic Cluster Expansion (ACE) framework \cite{Drautz2019, Drautz2020, Dusson2022}. We have used a julia implementation of the descriptor and fitting process and production simulations were performed using the Performant ACE evaluator\cite{Lysogorskiy2021} in the Large-scale Atomic/Molecular Massively Parallel Simulator (LAMMPS)\cite{LAMMPS}. The ACE1 julia code is under development at \url{https://github.com/ACEsuit/ACE1.jl}, and the exact version of ACE1 used in this project, including the modifications described below, can be found at \url{https://github.com/WillBaldwin0/ACE1.jl/tree/ortho-specdep-pairpot}. 

ACE models allow for a body-ordered expansion of the potential energy surface (PES) and the model presented in this work uses four body descriptors.

\subsubsection*{Two-body radial basis functions}

In this project, additional work was put into designing ACE models which robustly predicted strong interatomic repulsion at close approaches. In an ACE model, the distance between a pair of atoms is described by a set of radial basis functions\cite{Drautz2019}. In the ACE1 Julia implementation, this radial basis is defined as set of polynomials $p_n(x)$ which are orthogonal with respect to a weighting function $w(x)$:
\begin{align*}
    \delta_{nm} = \int_0^{r_{cut}} p_n(r) p_m(r) w(r) dr
\end{align*}
It is also possible to first transform the radial coordinate by a function $f$, so that $p(r) \rightarrow p(f(r))$, in order to provide more radial resolution at certain distances\cite{Kovacs2021}. For this project, a special choice was made for the weighting function $w(r)$ to encourage strong interatomic repulsion between atoms, even at distances much smaller than those seen in the training set. The weighting function $w(r)$ was constant beyond the typical nearest neighbour distance, but below this, it was shaped to approximately match the species dependent radial distribution function of the training set. The intuition behind this choice is that the weighting function is related to the regularisation of the potential energy landscape through the $L^2$ regularisation of the fitting procedure. If $w(r)$ is relatively large in some interval $r \in [a, b]$, the resulting potential energy landscape is heavily regularised in this region. By choosing $w(r)$ to match the radial distribution function, the potential energy landscape gradually becomes less regularised as the data becomes less frequent at small interatomic separations. This allows PES to be larger and have high curvature at small separations. This procedure was useful to ensure stable dynamics, but the shape of the weighting function did not affect the accuracy of the model.

\subsubsection*{Bayesian Regression and Active Learning}

As originally described by Drautz\cite{Drautz2019}, we use the atomic cluster expansion to construct a linear model of the PES. We subsequently fit the coefficients using Bayesian linear regression. The Bayesian linear regression algorithm returns the fitted parameters - which can be interpreted as the mean of the posterior distribution over the parameters - and a covariance matrix encoding the uncertainty in the solution. This covariance matrix can be used to infer an uncertainty in a future prediction. In practice, this is done by drawing a small number samples from the posterior parameter distribution, and constructing an ensemble of models.

The training procedure in this study involved two steps. Firstly, the Hyper Active Learning (HAL) Bayesian active learning framework \cite{vanderOord2022} was used to produce a model which was able to simulate the material over long time scales, with usable accuracy. A summary of the HAL procedure is as follows:
\begin{enumerate}
\item Generate a small initial database of training configurations labelled with DFT, for instance by perturbing an equilibrium structure. 
\item Fit an ACE model to this database using Bayesian linear regression. 
\item Run a molecular dynamics simulation using the ACE model, tracking the predicted uncertainty at each simulation step (see above). The dynamics is propagated based on a weighted sum of the potential energy, and a term representing the uncertainty of the models prediction. Hence, the dynamics is driven to areas in configuration space with a higher model uncertainty\cite{vanderOord2022}. 
\item If the predicted uncertainty exceeds some predetermined threshold, stop the simulation and perform the reference calculation on the current configuration.
\item Add this labelled configuration to the dataset, retrain the model, and restart the simulation.
\end{enumerate}
Once HAL had been used to create a stable model for the system, the resulting model was used to sample 214 configurations from a range of temperatures representative of the four phases of interest, which made up the final dataset. This included configurations of up to 160 atoms, sampled from both constant pressure and constant volume simulations. 

\subsubsection*{Reference Data}

All density functional theory calculations were performed using the FHI-aims DFT code at the LDA level of theory \cite{Blum2009}. A k-point density of 16 \AA$^{-1}$ was used, and `tight' basis set and integration grids were used throughout. 

\subsubsection*{Simulation Parameters}

Unless otherwise stated, all production simulations were performed in the constant pressure, constant temperature ensemble using a simulation cell of 69120 atoms, corresponding to $24^3$ pseudo cubic unit cells. The timestep was 4 fs. Statistics were collected by starting from the expected structure at the target temperature (obtained by sampling from slow cooling simulations) and equilibrating for 1 ns. A further nanosecond was then sampled every 200 fs. 

\subsubsection*{MYP Force Field Parameters}

\ce{CsPbI3} parameters correspond to Ref.\cite{Giri2022} but for a 7\% increase of the Cs-I Buckingham prefactor, here set to 150000 kcal mol$^{-1}$, in order to better reproduce the experimental transition temperatures.

\medskip
\textbf{Supporting Information}

Supporting Information will be made available on publication.



\bibliographystyle{unsrt}
\bibliography{bib_new}

\end{document}


\title{Dynamic Local Structure in Caesium Lead Iodide: Spatial Correlation and Transient Domains: Supplementary Information}
\maketitle

\section{Validation of Machine Learning Potential}

\subsection{Single-Point Accuracy and Efficiency}

We report the accuracy of the ACE model against the density functional theory baseline. Cesium lead iodide exhibits 4 distinct crystalographic phases between room temperature and 650K ~\cite{}, referred to as $\alpha$, $\beta$, $\gamma$ and $\delta$. The yellow, edge sharing $\delta$ phase is believed to be stable at room temperature. On cooling from 650K, where the material adopts the cubic $\alpha$-phase, the material transitions first to the tetragonal $\beta$, and then to orthorhombic, corner sharing $\gamma$ phase, before turning back into the $\delta$ phase once left at room temperature for a few seconds. Table \ref{table:point_accuracy} shows the accuracy of this model across unseen configurations of the four phases. These testing configurations consist of 160 atom unit cells drawn from constant temperature and pressure molecular dynamics performed using the final model.

\begin{table}[htb]
\caption{Single point accuracy of the ACE model quantified by Root mean square prediction errors}
\centering
\begin{minipage}{\linewidth}
\begin{tabular}{cccc} \toprule
\multicolumn{1}{p{1.9cm}}{\centering phase} & \multicolumn{1}{p{1.9cm}}{\centering Energy (meV/atom) } & \multicolumn{1}{p{1.9cm}}{\centering Force (eV/\AA) } & \multicolumn{1}{p{1.9cm}}{\centering Virial Stress (meV) } \\ \hline
$\delta$ & 0.780 & 0.032 & 4.504 \\
$\gamma$ & 0.619 & 0.026 & 4.351 \\
$\beta$ & 0.703 & 0.029 & 4.713 \\
$\alpha$ & 0.248 & 0.037 & 4.307 \\ \botrule
\end{tabular}
\end{minipage}
\label{table:point_accuracy}
\end{table}

\subsection{Phase Transition Temperatures}

The phase transition temperatures of this model are given in table \ref{table:transition_temps}. These values were computed by averaging the temperatures obtained from three heating and three cooling simulations at constant pressure with a cooling rate of 62.5 K/ns.

\begin{table}[htb]
\centering
\caption{Phase transition temperatures. The uncertainties correspond to one standard deviation across both heating and cooling simulations.}
\begin{tabular}{cc}
\toprule
Transition & Temperature (K) \\ \hline
$\gamma/\beta$       &  441 $\pm$ 8K  \\
$\beta/\alpha$      &  533 $\pm$ 9K   \\ \botrule
\end{tabular}
\label{table:transition_temps}
\end{table}

\section{Computing Octahedral Tilting Angles}

To assign three tilting angles to each octahedron, a regular octahedron is fitted to the \ce{PbI6} distorted octahedron as follows: Firstly, a right handed coordinate system aligned with the octahedra is formed by taking the body diagonals of the octahedron. This gives three unit vectors, which are aligned with the $x$, $y$, and $z$ axes of the octahedron (approximately aligned with the laboratory coordinate axes for small perturbations). These vectors can be considered as a matrix, which would be an orthogonal matrix if the vectors were perpendicular to one another. Secondly, the closest orthogonal matrix to this matrix is then found. This exactly orthogonal matrix can be interpreted as the rotation matrix mapping the laboratory coordinate axes onto a set of axes aligned with octahedron’s diagonals. The Euler angles can then be extracted from this rotation matrix. 

The intrinsic Euler angles were used in the order $x-y-z$. This means that for small rotations the three angles correspond approximately to rotations around the three coordinate axis. 

\section{Time Averaging and Low Pass Filters}

Many results reported in this work use a time average of the octahedral tilts. A square window was used to compute the average: Each output value is an average of 2 ps of the trajectory. We also recomputed several results using other methods to filter out high frequency components, to ensure that the observed features were not an artifact of the averaging process. In particular, we compared using a Butterworth low pass filter, with a critical frequency of $f = T^{-1}$ to averaging over a time $T$ of the trajectory.

Applying a first order Butterworth low pass filter to a signal corresponds to multiplying the Fourier transform of the signal by the following transfer function.
\begin{equation*}
	G(\omega) = \frac{1}{\sqrt{1+(\omega/\omega_{cut})^2}}
\end{equation*}
One can see that this suppresses the Fourier components of the signal with angular frequency higher than $\omega_{cut}$, and leaves the components with angular frequency lower than $\omega_{cut}$ largely unaffected.

Using a filter of this kind gave very similar results to the square window rolling average. 

\section{Fits to Autocorrelation Functions}

\begin{figure}
    \centering
    \includegraphics[width=0.7\linewidth]{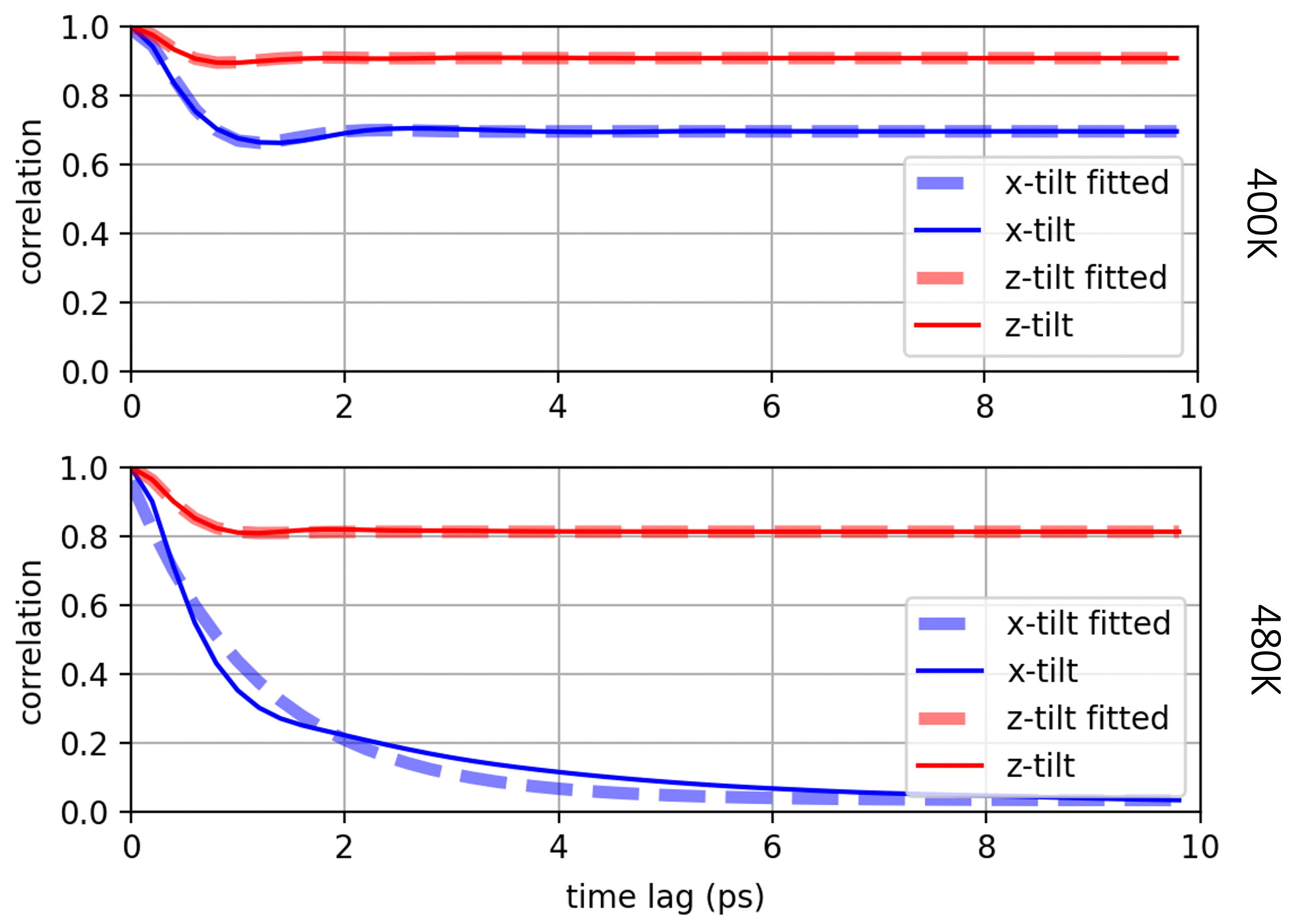}
    \caption{Temporal autocorrelation of octahedral tilts fitted to the decay of a damped harmonic oscillator. In Both the orthorhombic and cubic phases, the tilt around the $y$-axis is equal to that around the $x$-axis, and has therefore been omitted.}
    \label{fig:corr_fits}
\end{figure}

The temporal autocorrelation functions of individual octahedral tilts can be fitted to the decay of a damped harmonic oscillator \cite{Fransson2022}. The result of this process is shown in Figure \ref{fig:corr_fits}. For all the ordered tilts (where the correlation does not approach zero) the fit is quite good, while for the disordered $z$-direction tilt in the tetragonal tilt, this model does not capture the shape of the decay.

\section{Assessing independence of different tilting angles}
\label{sec:independent_tilts}

Once can assess how coupled the distributions of $x$- and $y$-direction tilt by comparing the true joint distribution to the distribution assuming the two variables are independent. Let $X$ and $Y$ denote the two tilting angles, and write the measured joint distribution as \begin{align*}
    P(X,Y)
\end{align*}
If $X$ and $Y$ are independent, then
\begin{align*}
    P_{\text{independent}}(X,Y) = P(X)P(Y)
\end{align*}
where
\begin{align*}
    P(X) = \int P(X,Y=y)dy 
\end{align*}
We can therefore compute and $P_{\text{independent}}(X,Y)$ and compare with $P(X,Y)$.

\section{Temperature Dependence of 2D and 3D Tilting Angle Distributions}

\begin{figure}
    \centering
    \includegraphics[width=0.95\linewidth]{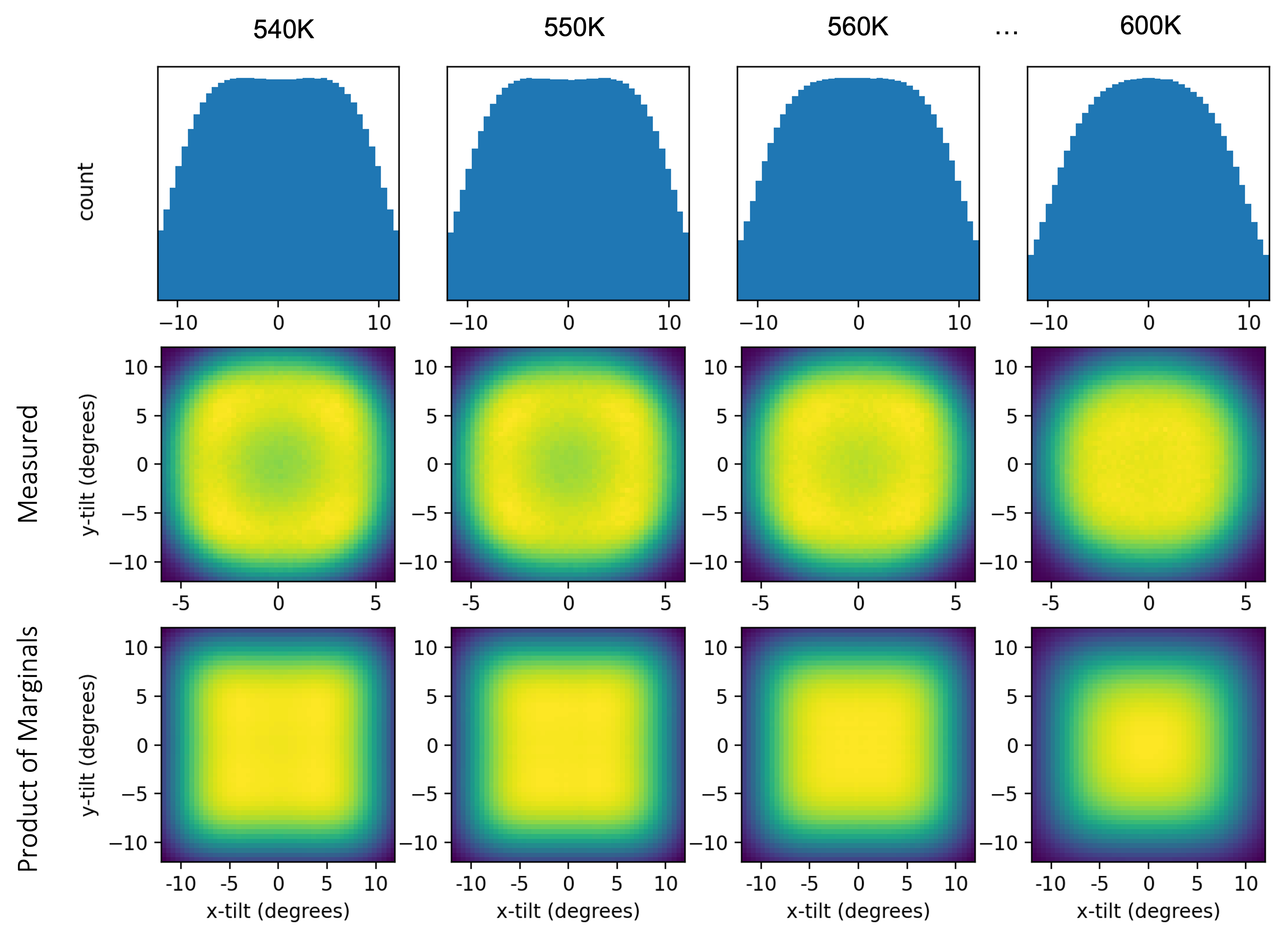}
    \caption{Tilting angle distributions as a function of temperature. Top row: histograms of $x$-direction tilt angle. Middle row: observed joint distribution of $x$- and $y$- tilt. Bottom row: Product of marginal distributions for $x$- and $y$- tilts.}
    \label{fig:temperature_dep_heatmaps_2ps}
\end{figure}

The time averaged single and joint distributions of tilting angles show some temperature dependence. Figure \ref{fig:temperature_dep_heatmaps_2ps} shows this variation in the cubic phase close to cubic to the tetragonal transition temperature of 533 K. The single tilt distribution, as well as the two tilt joint distribution and the product of the marginal distributions is shown. These figures correspond to a 2 ps square window rolling average. Once can see that for the 2D joint distribution, the multimodality is lost at 600 K.

\begin{figure}
    \centering
    \includegraphics[width=0.5\linewidth]{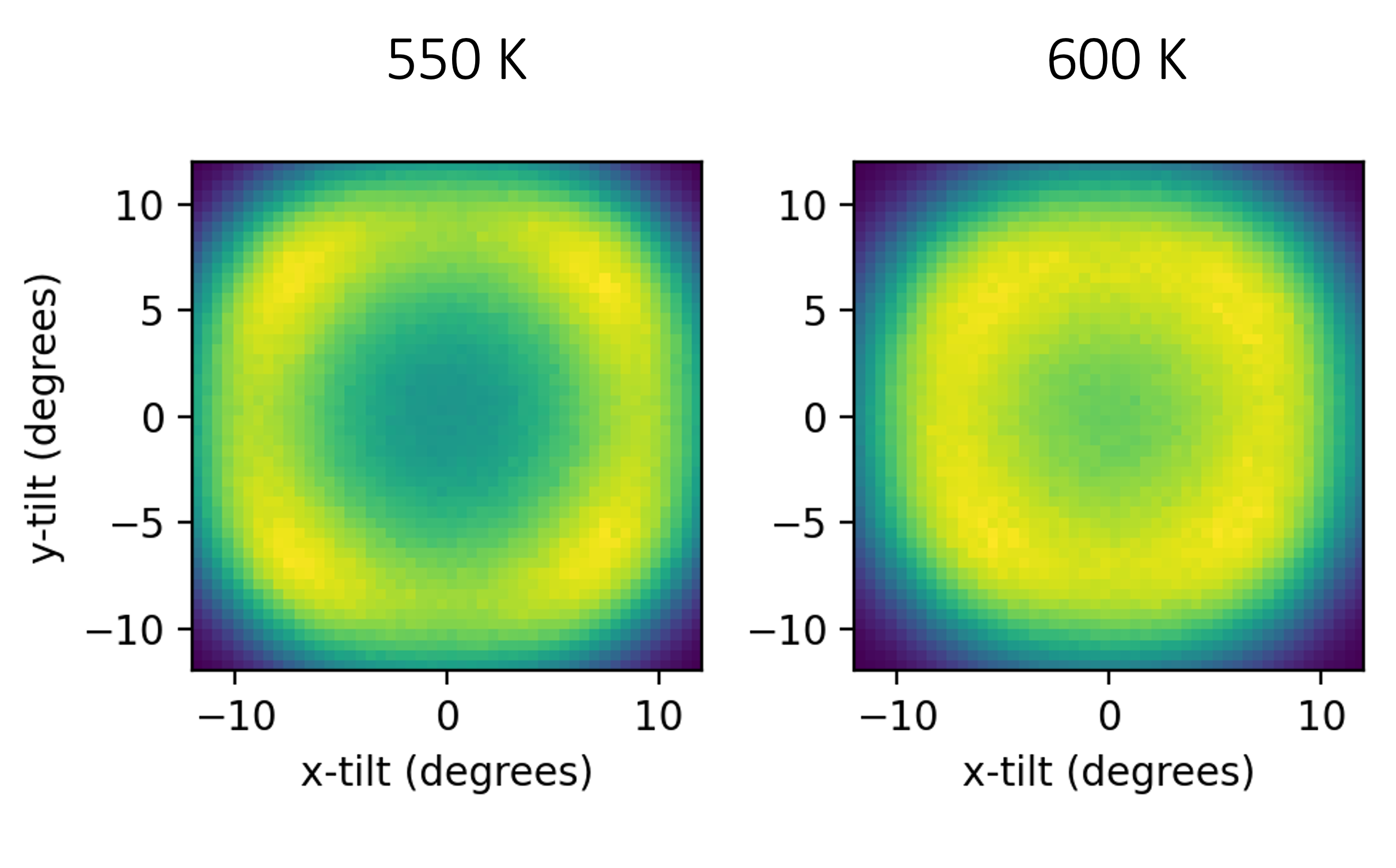}
    \caption{A slice through the 3D tilting angle joint distribution around $\theta_z=0$. These plots show the tilts after applying 2 ps rolling average.}
    \label{fig:3d_slices}
\end{figure}

The tilting angle vector $(\theta_x, \theta_y, \theta_z)$ has distribution over $\mathbb{R}^3$. The 2D joint distributions shown in figure \ref{fig:temperature_dep_heatmaps_2ps} correspond to integrating over one component of this distribution:
\begin{equation*}
    P_{XY}(\Theta_x, \Theta_y) = \int P(\Theta_x, \Theta_y, \Theta_z=\theta_z) d\theta_z
\end{equation*}
It is also possible to visualise the full 3D distribution by plotting slices through it at a fixed value of $\theta_z$. This reveals that the time averaged distribution of tilts remains multimodal far above the transition temperature. In Figure \ref{fig:3d_slices}, the distribution of $(\theta_x, \theta_y)$ for a slice around $\theta_z=0$ is plotted for a range of temperatures. Specifically, we have taken all points where $-4 < \theta_z < 4$ (with angles measured in degrees):
\begin{equation*}
    P_{XY,Z\sim 0}(\Theta_x, \Theta_y) := \int_{-4}^{4} P(\Theta_x, \Theta_y, \Theta_z=\theta_z) d\theta_z
\end{equation*}
Here we find that at 600K, 77 K above the transition temperature of this potential, there is still a strong tendency to avoid having all three tilting angles equal to zero. 

\section{Videos}

Supplementary videos 1 and 2 show the time variation of the spatial planar regions of correlated tilting angle at 450 K and 550 K respectively over a 400 ps trajectory. High resolution versions are available at \url{https://www.dropbox.com/sh/g2f97ajfse8u340/AADuaOru7MZBx1CsuzNLBN0ha?dl=0}.

In supplementary video 1, the three tilting angles $\theta_x, \theta_y$ and $\theta_z$ are visualised from left to right. In video 2, only $\theta_x$ and $\theta_y$ are shown. As in the main text Figure 4b and Figure 5b, the time averaged tilting angles are first masked such that homogeneous regions are rendered the same colour, as discussed in main text sections 2.4 and 2.5. 

These videos show the time averaged tilting angles, where the average was taken over 2 ps. Equivalent videos without first performing the time average have a similar visual appearance. 

\bibliography{bib_new}